\documentclass[lettersize,journal]{IEEEtran}
\usepackage{amsmath,amsfonts}
\usepackage{array}
\usepackage[caption=false,font=small,labelfont=sf,textfont=sf]{subfig}
\usepackage{textcomp}
\usepackage{stfloats}
\usepackage{url}
\usepackage{verbatim}
\usepackage{graphicx}
\usepackage{cite}
\usepackage[ruled]{algorithm2e}

\usepackage{color}
\usepackage{amssymb}
\usepackage{booktabs}
\usepackage{comment}
\usepackage{bm}
\begin{document}

\title{{\sffamily Egret}: Reinforcement Mechanism for Sequential Computation Offloading in Edge Computing}

\author{Haosong~Peng,
         Yufeng~Zhan,
         Di-Hua~Zhai,
         Xiaopu~Zhang,
         and~Yuanqing~Xia,~\IEEEmembership{Fellow,~IEEE}
 \IEEEcompsocitemizethanks{\IEEEcompsocthanksitem H.~Peng, Y.~Zhan, D.-H.~Zhai, and Y.~Xia are with the School
 of Automation, Beijing Institute of Technology, Beijing,
 China, 100086.
 E-mail: livion\_i@icloud.com~(Peng),~yu-feng.zhan@bit.edu.cn~(Zhan), zhaidih@bit.edu.cn~(Zhai), xia\_yuanqing@bit.edu.cn~(Xia)
 \IEEEcompsocthanksitem X.~Zhang is with iF-Labs, Beijing Teleinfo Technology Co., Ltd., Beijing, China.
 E-mail: zhangxiaopu@teleinfo.cn
 }%
 \thanks{This work has been submitted to the IEEE for possible publication. Copyright may be transferred without notice.}
 }

\markboth{Journal of \LaTeX\ Class Files,~Vol.~14, No.~8, August~2021}%
{Shell \MakeLowercase{\textit{et al.}}: A Sample Article Using IEEEtran.cls for IEEE Journals}

\maketitle

\begin{abstract}
As an emerging computing paradigm, edge computing offers computing resources closer to the data sources, helping to improve the service quality of many real-time applications. 
A crucial problem is designing a rational pricing mechanism to maximize the revenue of the edge computing service provider~(ECSP).
However, prior works have considerable limitations: clients are static and are required to disclose their preferences, which is impractical in reality.
To address this issue, we propose a novel sequential computation offloading mechanism, where the ECSP posts prices of computing resources with different configurations to clients in turn. 
Clients independently choose which computing resources to purchase and how to offload based on their prices. 
Then Egret, a deep reinforcement learning-based approach that achieves maximum revenue, is proposed. 
Egret determines the optimal price and visiting orders online without considering clients' preferences. 
Experimental results show that the revenue of ECSP in Egret is only 1.29\% lower than Oracle and 23.43\% better than the state-of-the-art when the client arrives dynamically.

\end{abstract}


\section{Introduction}\label{sec:introduction}

\IEEEPARstart{R}eal-world applications such as autonomous driving~\cite{10.1145/3447993.3483242,10077757}, video analytics~\cite{10.1145/3372224.3380881,9953098}, and AR/VR~\cite{10.1145/3300061.3300116,10033423}, the immediate processing of computing tasks needs to be guaranteed. 
Although computing tasks can be transferred to the cloud, where resources are relatively sufficient, the substantial transmission overhead makes it burdensome for edge devices.
Thanks to edge computing, a merging paradigm that situates computation resources closer to data, this adjacency not only reduces transmission overhead but also enhances the efficiency and responsiveness of applications, ensuring these latency-sensitive applications operate seamlessly in real-time. 

In a typical edge computing scenario, an edge computing service provider (ECSP) is the one who offers computation resources (i.e., VM, container, etc.) as commodities, catering to the immediate processing services of various applications for edge devices. 
Due to the limited computational resources and battery life of edge devices, a common practice is to offload partial~\cite{10.1145/3384419.3430898,9847073} or entire~\cite{Khani_2021_ICCV,276952} computation-intensive tasks to the ECSP, thus reducing latency and energy consumption. 
In return, edge devices are required to pay the fee for the corresponding resources. 
For the ECSP, the challenge is how to price these resources to maximize its revenue while satisfying the utility of the edge device.

Previous methods~\cite{li2021deep,8744396,9040268} have explored how to maximize the revenue of ECSP. 
However, they deviated from the conditions in real-world scenarios.
First, the clients are unwilling to disclose their private information (i.e., data size, local computing capacity, etc.). 
Instead, they are more inclined to select resource configurations that suit their needs. 
Second, the number of clients varies, and tasks arrive dynamically at the system. 
Therefore, the future arrival of clients is unknown, and even the availability of ECSP resources is unknown.
In this situation, offloading decisions and pricing decisions must be made in an online manner.
Finally, the group strategyproofness~\cite{goldberg2005collusion} is hard to ensure, implying that multiple clients may collaborate to manipulate the mechanism to work against the server.

To address these challenges, we propose a novel sequential computation offloading mechanism (SCOM) in this paper.
In this mechanism, the ECSP owns computing resources with different configurations. 
It posts prices to the clients in a predetermined visiting order, wherein each client independently determines which configurations to purchase and the number of tasks to offload according to the posted prices.
The ECSP visits the next client and repeats the above process until all clients have been visited or all the resources are sold out.
This more realistic mechanism brings many advantages to the computation offloading problem.
First, the 'take-it-or-leave-it' rule naturally makes it a group strategyproof mechanism~\cite{goldberg2005collusion}. 
Second, clients do not need to report their preferences precisely. 
They adopt a dominant strategy in response, driven by their utilities.
Therefore, there's no risk of exposing their private information.
Finally, clients can determine whether they offload or not immediately.

To maximize the revenue of the ECSP, we design Egret, an experience-driven method, allowing the ECSP to learn pricing strategy and bidding order from historical transaction records.
Deep Reinforcement Learning (DRL) is a machine learning paradigm similar to human learning from experience.
It's a reward-oriented learning method that involves interaction with the environment. 
Specifically, the agent takes action based on observation, and then the environment returns a reward to stimulate the agent's strategy updates. 
By continuously interacting with the environment, the agent can learn which action to take to maximize its reward. 
In Egret, we only need to utilize the clients' labels and historical transaction records to ensure privacy.
Besides, the online decision-making feature of Egret makes it suitable for scenarios in which clients arrive dynamically.
Overall, the contributions of this paper are summarized as

\begin{enumerate}

\item We propose a novel sequential computation offloading mechanism and study how to maximize the revenue of ECSP in static and dynamic settings.
We give the theoretical optimum that determines the pricing strategy and the client visiting order in an Oracle scenario where the client preference information and the remaining resources of the ECSP are known.

\item We design Egret, an DRL-based approach that achieves near-optimal revenue for the ECSP only depending on the experience in an online manner.

\item Experimental results show that Egret outperforms other baselines. Egret's optimal result in the SCOM setting is only 1.289\% lower than Oracle and 23.43\% better than the state-of-the-art in a dynamic SCOM setting.
\end{enumerate}

The rest of this paper is organized as follows. 
Section~\ref{sec_2} describes the system model and establishes the problem formulation.
Section~\ref{sec_3} proposes the design of Egret. 
Evaluations are presented and discussed in Section~\ref{sec_4}. 
Section~\ref{sec_5} introduces the related works. 
Finally, Section~\ref{sec_6} concludes the paper.

\section{System Model and Problem Analysis}\label{sec_2}
In this section, we first introduce the economic framework of the sequential price mechanism (SPM).
Then, We establish the system model of computation offloading consisting of two types of computational models and the utility function of the client and formulate this problem as a SCOM. 
Next, we provide the solution of optimal strategy when we know the necessary information. 
Last, we extend the SCOM to the dynamic SCOM in more realistic scenarios where clients dynamically enter and exit.

\subsection{Sequential Price Mechanism}

In an SPM context\cite{goldberg2005collusion,chawla2010multi}, there are $n$ buyers and $m$ items for sale. 
The seller engages in a sequential interaction with each buyer. 
At the beginning of each round, the seller selects an unvisited buyer and proposes prices for all remaining items that have not yet been sold. 
The buyer selects the item that maximizes its utility. 
The seller records the transaction, including the buyer, the purchased item, and the payment. 
This process repeats until all items have been sold or all buyers have been visited. 
The objective of the seller is to maximize its revenue. 

\begin{table}[!t]
\setlength{\abovecaptionskip}{0cm}  %
\setlength{\belowcaptionskip}{0cm} %
\renewcommand{\arraystretch}{1.1}
\centering
\caption{Notation and Descriptions}
\label{tab1}
\setlength{\tabcolsep}{8mm}{
\resizebox{\columnwidth}{!}{%
\begin{tabular}{c|c}
\toprule
\textbf{Notation} & \textbf{Description}                        \\ \midrule
$\mathcal N$                 & set of clients                     \\
$\mathcal M$                 & set of instances                   \\
$d_i$                 & the input data of client $i$              \\
$x_{ij}$                 & data from client $i$ offloaded to instance $j$ \\
$f_{i}$                & computing capacity of client $i$            \\
$b_{i}$                 & average network bandwidth of client $i$                  \\
$\mu_{i}$                & energy dissipation coefficient of client $i$             \\
$\nu_{i}$                 & transmission cost coefficient of client $i$               \\
$\tau_{ij}$               &payment of client $i$ to rent instance~$j$         \\
$p_{j}$                 & price of instance $j$                        \\
$\alpha,\beta,\gamma$              & hyper-parameter of client's utility      \\ \bottomrule
\end{tabular}%
}}
\end{table}

\begin{figure}[!b]
\begin{center}
\setlength{\abovecaptionskip}{0.cm}
\includegraphics[width=0.9\linewidth]{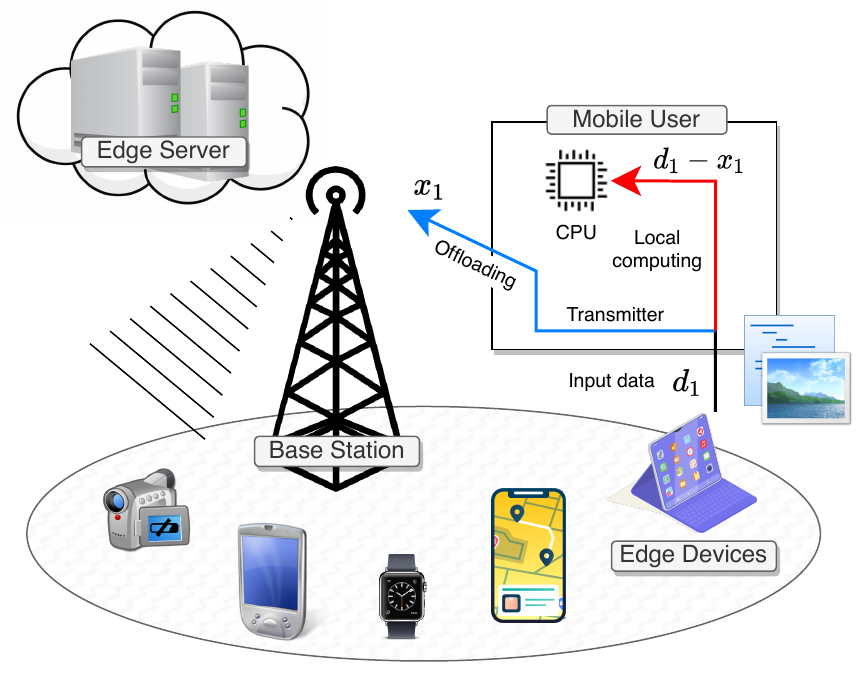}
\caption{Multi-user computation offloading in an edge computing environment.}
\label{fig:1}
\end{center}
\end{figure}

\subsection{System Model}
As shown in Fig.~\ref{fig:1}, there are $N$ clients, and the ECSP owns $M$ resource instances with different configurations. 
Let $\mathcal N = \{1, . . . , N\}$ be the set of clients and each of which has a computing task to be executed, where~$d_i$ denotes the input data of client~$i$. 
The client can choose to compute the task locally or complete a part of the task in the edge server.
Let $x_{ij}$~$(0\leq x_{ij}\leq d_i)$ be the decision of client~$i$, denoting the data that client~$i$ offloads to the instance~$j$ in the edge server, hence~$d_i-x_{ij}$ is the data remains at local. 

Each client is a mobile user belonging to a distinctive type of model whose real preferences and characteristics (i.e., bandwidth, computing capacity, etc.) are inaccessible to the ECSP because of privacy. 
The average network bandwidth and computing capacity assigned to client $i$ are denoted by $b_{i}$ and $f_{i}$, respectively.
The ECSP has a set of configured instances $\mathcal M=\{1, 2,\cdots, M \}$, where the computing capacity of instance~$j$ is~$F_j$, and~$F_j < F_{j+1}$.
In this paper, we set a higher price for the instance with higher computing capacity, i.e., the price of instance~$j+1$ is greater than~$j$. 
All the symbols mentioned in Sec.~\ref{sec_3} are listed in Table~\ref{tab1}.

\textbf{Local computing:}
Client~$i$ offloads a part of its computation task~$x_{ij}$ to instance $j$ in the edge server. Then, the latency of local computing is~\cite{tran2019federated,zhan2020deep}
\begin{equation}
t_{i}^{loc}= \frac{d_{i}-x_{ij}}{f_{i}}.
\end{equation}
According to the widely used energy model, the energy cost of client~$i$ in the local computing is~\cite{tran2019federated,burd1996processor,zhan2021l4l}
\begin{equation}
e_{i}^{loc} = \mu_{i}(d_{i}-x_{i})f_{i}^{2},
\end{equation}
where~$\mu_{i}$ is the energy dissipation coefficient.  

\textbf{Computation offloading:}
The offloading latency required by client~$i$ to process the computation task remotely includes offloading time and execution time on the edge server. 
Clients select the instance according to the ECSP's pricing strategy. Assuming client $i$ selects instance $j$ to execute its task remotely, then the offloading latency can be defined as
\begin{equation}
t_{ij}^{off}= \frac{x_{ij}}{b_{i}} + \frac{x_{ij}}{F_{j}},
\end{equation}
where~$b_i$ is the network bandwidth assigned to client~$i$.
The transmission cost of client~$i$, depends on its transmission bandwidth $b_{i}$ and offloading data size $x_{ij}$, is~\cite{zhan2020deep,burguera2011crowdroid,xiao2017cloud} 
\begin{equation}
e_{i}^{off} = \nu_{i}\frac{x_{ij}}{b_{i}},
\end{equation}
where $\nu_{i}$ is the transmission cost coefficient.
In addition, based on the cloud computing on-demand instance pricing~(e.g., AWS EC2 on-demand instance pricing~\cite{awsec2}), the instance procurement cost of client~$i$ is
\begin{equation}
\label{tau}
\tau_{ij} = p_{j}\frac{x_{ij}}{F_{j}},
\end{equation}
where $p_{j}$ is the price set by ECSP for instance $j$.

\textbf{Client Utility Model:}
As shown in Fig.~\ref{fig:1}, the computation task of client $i$ is split into two parts, the local computation and remote computation, which can be executed simultaneously. 
Therefore, the time latency to complete the task of client $i$ is 
\begin{equation}
t_{i} = \max\{t_{i}^{loc}, t_{ij}^{off}\} = \max\{\frac{d_{i}-x_{ij}}{f_{i}},\frac{x_{ij}}{b_{i}} + \frac{x_{ij}}{F_{j}}\}.
\end{equation}
The total energy cost of client $i$ includes local computing and offloading transmission, which is
\begin{equation}
e_{i} = e_{i}^{loc} + e_{i}^{off} = \mu_{i}(d_{i}-x_{ij})f_{i}^{2} + \nu_{i}\frac{x_{ij}}{b_{i}}.
\end{equation}
Based on the energy cost, the execution latency, and the additional instance procurement cost of client $i$, the utility function of client $i$ can be deﬁned as
\begin{equation}
\label{eq:u0}
\begin{aligned}
U_{i}\left(p_{j}, x_{ij},F_{j}\right)=&\alpha e_{i}+\beta t_{i}+\gamma \tau_{ij}\\
 =& \alpha \left[\mu_{i}(d_{i}-x_{ij})f_{i}^{2} + \nu_{i}\frac{x_{ij}}{b_{i}}\right]\\ &+ \beta \left[\max\{\frac{d_{i}-x_{ij}}{f_{i}},\frac{x_{ij}}{b_{i}} + \frac{x_{ij}}{F_{j}}\}\right]
\\ &+ \gamma \left[p_{j}\frac{x_{ij}}{F_{j}}\right],
\end{aligned}
\end{equation}
where $\alpha$, $\beta$, and $\gamma$ are the adjustable parameters to provide rich modeling flexibility, and the parameters represent the clients' preferences. For example, the clients pay more attention to energy costs with larger $\alpha$. 
Each client aims to determine the optimal instance type and offloading strategy to minimize the utility.

\subsection{Problem Formulation}\label{sec:pf}
When clients have computation offloading requirements, the ECSP will visit them round by round.
At each round, the ECSP visits client~$i$, then set the prices $\mathbf P^k = (p_{i1}^k,p_{i2}^k,\cdots,p_{iM}^k)^T$ for the remaining instances according to its policy. 
If instance $j$ has been rented, client~$i$ cannot rent it so that $p_{ij}^k=+\infty$.
After receiving the prices, aiming to minimize the utility, the client independently decides which type of instance to rent and how to offload. 
The ECSP records the payment, the objective of which is to maximize the revenue.

At round $k$, the ECSP maintains the following information: 
1) a set of residual clients waiting to be visited and a set of residual instances that have not been rented,
2) a temporary matrix $\mathbf X^k \in \mathbb R^{N\times M}$ records the purchase trace, which encodes $\mathbf X^{k}[i][j] = 1$ when client~$i$ selects instance $j$, otherwise, $\mathbf X^{k}[i][j] = 0$, 
3) a payment matrix $\mathbf{Y}^k\in \mathbb{R}^{N\times M}$, similar to $\mathbf{X}^k$, that stores the payment $\tau_{ij}^k$.
For example, if client $i$ purchase the resourece $j$ at payment $\tau_{ij}^k$, $\mathbf{Y}^k[i][j]$ is set to $\tau_{ij}$.

The SCOM starts by initializing~$\mathbf X^{1} = \mathbf 0$,~$\mathbf Y^{1} = \mathbf 0$ updates all the information at the end of each round, and terminates when all the clients have been visited, or all the resources have been rented. 
The goal of the ECSP is to maximize the expected revenue, then the optimized problem of the ECSP can be defined as
\begin{subequations}
\begin{align}
\max &\quad \sum_{k=1} \mathbf {Y}^k,  \label{op1}\\
s.t. &\quad \sum_{i=1}^{N} \mathbf X^k[i][j] \le 1, \quad j\in \mathcal{M}, \label{eq:st1} \\
\quad &\quad \sum_{j=1}^{M} \mathbf X^k[i][j] \le 1, \quad i \in \mathcal{N}, \label{eq:st2} \\
\quad  &\quad \mathbf X^k[i][j]\in\{0,1\}, \label{eq:st3}
\end{align}
\end{subequations}
where Eq.~(\ref{eq:st1}) indicates that every client will only rent one instance, and Eq.~(\ref{eq:st2}) ensures that every instance can only be rented by one client at the same time.

\subsection{Optimal Strategy with the Oracle}\label{opt}
Each client's private parameters, such as bandwidth $b_i$ and computing capacity $f_i$, etc., are observable in an Oracle scenario.  
It is straightforward to obtain the best visiting order and pricing strategy if the ECSP knows the complete utility information of each client. 
From the perspective of a client, its objective is to minimize the utility, which can be depicted as
\begin{equation}
\label{eq:u}
\begin{aligned}
\min \ & \quad U_i\left(p_j, x_{ij},F_j\right)= \alpha e_i+\beta t_i + \gamma \tau_{ij},\\
s.t.& \quad 0 \leq x_{ij} \leq d_i.
\end{aligned}
\end{equation}
Noticing that the price $p_j$ and resources configuration $F_j$ are given by the ECSP. 
According to Eqn.~(\ref{eq:u0}), the optimization problem~(\ref{eq:u}) can be transformed into two piecewise optimization problems, as

\begin{equation}
\label{eq:u1.2}
\begin{aligned}
\min \ & \quad U_{i,1}=(-\alpha\mu_i f_i^{2}+\frac{\alpha \nu_i}{b_i}-\frac{\beta}{f_i}+\frac{\gamma p_j}{F_j})x_{ij} \\
& \quad +\alpha\mu_i d_i f_i^{2}+\frac{\beta d_i}{f_i},\\
s.t. & \quad 0 \leq x_{ij} \leq \frac{d_i b_i F_j }{b_i F_j +f_i F_j + f_ib_i}.
\end{aligned}
\end{equation}

\begin{equation}
\label{eq:u1.1}
\begin{aligned}
\min \ & \quad U_{i,2}=(-\alpha\mu f_i^{2}+\frac{\alpha \nu_i+\beta}{b_i}+\frac{\beta+\gamma p_j}{F_j})x_{ij} \\
& \quad + \alpha\mu_i d_i f_i^{2}, \\
s.t. \quad & \frac{d_i b_i F_j} {b_iF_j+f_iF_j+f_ib_i} \leq x_{ij} \leq d_i. 
\end{aligned}
\end{equation}
Noting that $U_i$ is a piecewise linear function, the $U_i$ achieves the minimum value when $x_{ij}$ equals to $0$, $\frac{d_ib_iF_j}{b_iF_j+f_iF_j+f_ib_i}$ or $d_i$. Setting $U_{i,ini}=U_i(p_j,0,F_j)$, $U_{i,mid}=U_i(p_j,\frac{d_ib_iF_j}{b_iF_j+f_iF_j+f_ib_i}, F_j)$, and $U_{i,end}=U_i(p_j,d_i,F_j)$.

Since $U_{i,ini}$ does not depend on the $p_i$, which allows us to calculate the optimal price $p^{*}$ by solving two equations, as 
\begin{equation}
\begin{cases}
 U_{i,mid}=U_{i,ini}\rightarrow p_{a}, \\
 U_{i,end}=U_{i,ini}\rightarrow p_{b},
\end{cases}
\end{equation}

\begin{equation}\label{eq_2-4-5}
p^{*} = max(p_{a},p_{b}).
\end{equation}

Overall, once the ECSP offers the resource $F_j$ with a price slightly smaller than $p^{*}$, the client has a motivation to purchase the resource because it leads to a smaller utility.

Finally, we have to get the upper bound price that exactly persuades the client to purchase the resource. 
First, initialize the visiting trace and best prices (Lines 1-2).
Based on the analysis of the practical model,  we can obtain the maximum price $p^{*}_{i,j}$ of resource $j$ to client $i$ and compose all the prices as $\mathbf P_{i}^{*} = (p^{*}_{i1},p^{*}_{i2},\cdots,p^{*}_{iM})^T$.  
Further, we repeat this process for every awaiting client $\mathbb P = [\mathbf P_{1}^{*}, \mathbf P_{2}^{*},\cdots \mathbf P_{N}^{*}]^T$ (Lines 3-8). 
Last, the optimal visiting order and prices are easily yielded from the greedy approach (Lines 9-14). 

\begin{algorithm}[!t]  %
	\caption{Optimal Strategy with the Oracle}%
	\LinesNumbered %
	\KwIn{the utility function of clients $\mathbf U_i$, clients number~$N$ and resource number~$M$} %
	\KwOut{the optimal visiting order and prices}%
	trace  $\leftarrow[\ ]$        \#initialize the visiting trace $\in\mathbb R^{N}$\;
        prices  $\leftarrow[+\infty ]$	\#initialize the best prices $\in\mathbb R^{N\times M}$\;
	\For{$i=1,2,\ldots,N$}{
		\For{$j=1,2,\ldots, M$}{
		$\mathbb P[i][j]$ $\leftarrow$ \textit{get\_maximum\_price$(U_{i})$} according to Eqn.~(\ref{eq_2-4-5}) \;
		$\tau[i][j]$ $\leftarrow$ \textit{get\_payment$(\mathbb P[i][j])$} according to Eqn.~(\ref{tau})
		}
	}
	\For{$i=1,2,\ldots,\min(N,M)$}{
	row, col $\leftarrow$ \textit{where}($\bm\tau$, $\max \bm\tau$) \#find the position of the largest element in $\bm\tau$\;
        prices[row][col] $\leftarrow \mathbb P[row][col]$\;
	trace.\textit{append}(row)\;
	$\tau[row][\cdot] \leftarrow \mathbf 0$\;
	$\tau[\cdot][col] \leftarrow \mathbf 0$\;
	}
	\Return trace, prices
\end{algorithm}

\subsection{Dynamic Sequential Computation Offloading Mechanism}
The SCOM only supports situations where the set of clients is static, and the ECSP sells the resources once. 
However, occasionally, the clients come in batches at any time, and the ECSP recycles the allocated resources once the clients finish their tasks. We refer to it as the dynamic sequential computation offloading mechanism (DSCOM).

Similar to SCOM, considering there are $N$ different types of clients, the ECSP owns $M$ resources of different configurations. 
Clients can dynamically enter the edge computing environment in a batch at any time and wait for the bid from the ECSP. 
The set of clients entering simultaneously is arbitrary. 
At the end of each time interval, the ECSP makes the sequential decisions of visiting order and prices for the waiting clients at one time, i.e., they finish one SCOM. 
Then, the time shifts to the next time interval. 
Whenever one client completes its task, the resource is freed and recycled. 
Our goal is to maximize the total revenue of the ECSP. 

\section{Design of Egret}\label{sec_3}
In this section, we first introduce the overview of our system. 
Then, we propose Egret, an experience-driven method for solving the optimal SCOM problem.
The Markov decision process description and the RL training algorithm are thoroughly discussed.

It's non-trivial to solve the optimization problem~(\ref{op1})-(\ref{eq:st3}) mentioned in Sec.~\ref{sec:pf} because of two reasons. 
On the one hand, the ECSP has little information about the clients' parameters and configurations (e.g., $f_i$, $\nu_i$, etc.), as the clients are unwilling to expose their private preferences and for security reasons. 
Therefore, the ECSP cannot make the optimal pricing strategy if don't have the client's oracle information, such as utility (Sec.~\ref{opt}). 
Previous work~\cite{daskalakis2012optimal} points out that it is intractable to optimally price multiple items for a buyer whose utility is implicitly given by a closed-form formula, let alone more clients. 
On the other hand, although serval studies \cite{huo2022learning,brero2021reinforcement,brero2021learning} have focused on solving simplified SPM from a learning perspective, this problem is even more complicated in a dynamic environment. 
They may fail to capture their utilities when facing different unseen combinations of clients, resulting in poor performance. 

\begin{figure}[!t]
\begin{center}
\setlength{\abovecaptionskip}{0.cm}
\includegraphics[width=1.0\linewidth]{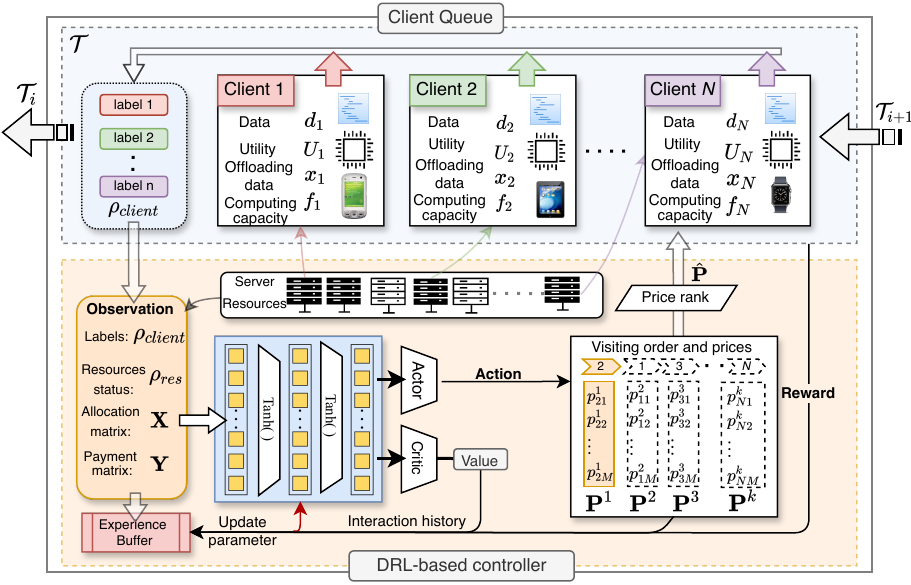}
\caption{The system overview.}
\label{fig:2}
\end{center}
\end{figure}

To address such challenges, we present an experience-driven method to achieve optimal overall revenue by deciding on resource pricing and visiting orders.
It is enlightened by the insights that the ECSP usually profiles clients based on their behavior and purchase history. 
Therefore, Egret categorizes clients into different labels to better respond when dealing with the same label client next time. 

\subsection{Overview} 
As shown in Fig \ref{fig:2}, at time interval $\mathcal T$, there are $n$ types of clients arrive in the \textit{client queue} with different utilities $U_i$ and local computing capacity $f_i$, each of which has its task data $d_i$ and best response strategy. 
The \textit{DRL-based controller} in the server observes the clients' labels from the queue, resources status, and other accessible information and feeds them to the policy network. 
The actor-network outputs the visiting order and prices of resources, and then the ranked prices are offered to the selected client. 
After that, the client calculates its utility and determines which type of resource $F_j$ to buy and how much data $x_{ij}$ to offload to the ECSP. 
Finally, a payment calculated according to the resource occupied time returns to the ECSP as the \textit{reward}.

\subsection{DRL Model}
\textbf{State Space.} The state $s^{k}=(\rho^{k},\mathbf X^{k},\mathbf Y^{k})$ consists of the remaining clients and resources $\rho^{k}_{client}$, $\rho^{k}_{res}$, the allocation matrix $\mathbf X^{k}$, and the payment matrix $\mathbf Y^{k}$.  
The $\rho^k$ is a residual set of clients and resources $\rho^{k} = (\rho^{k}_{client},\rho^{k}_{res})$, where $\rho^{k}_{client}\in \mathbb R^{N}$ and $\rho^{k}_{res}\in \mathbb R^{M}$ are the one-zero vectors encoding the status of clients and resources, respectively.
Noting that $\rho^{k}_{client}[i] = 1$ indicates the $i$-th client has not been visited by the ECSP and $\rho^{k}_{res}[j]=1$ indicates the instance $j$ has not been rented. 

The information provided to the ECSP plays a pivotal role in learning an optimal strategy. 
With the help of lemmas from~\cite{brero2021reinforcement}, we can elucidate the rationale behind our design of the statistic as the state space for Egret.

\textbf{Lemma 1.} \textit{For a sequential price mechanism, the statistic of the allocation matrix and the clients who have not been visited is sufficient to determine the optimal policy, whatever the design objective. Moreover, there exists a unit-demand setting with correlated clients' utility where the optimal policy must use a sufficient statistic of size $\Omega(\max \{n, m\} \log (\min \{n, m\}))$}

Lemma 1 offers insights into the minimum requirements of state space information. 
The remaining clients and resources are insufficient to develop an optimal policy since they belong to linear space. 
It's necessary to provide the allocation matrix and remaining clients as state space. 
However, following numerous experiments with state design, we believe that the payment matrix also constitutes practical knowledge for the agent, as it speeds up the convergence of the DRL algorithm.

\textbf{Action Space.} An action $a^{k}=(I^{k}, \mathbf P^{k})$ defines that the ECSP visits the $I^{k}$-th client and bid prices $\mathbf P^{k}$ at round $k$.

\textbf{Transition.} 
The state and action designs apply to both SCOM and DSCOM, but the state transition needs to be distinguished in detail.
For an SCOM, the selected client offloads its data $x_{ij}$ and chooses to buy a resource $F_j$ or not, leading to a new state by updating the information $\rho^{k}\rightarrow \rho^{k+1}$, $\mathbf X^{k}\rightarrow \mathbf X^{k+1}$ and $\mathbf Y^{k}\rightarrow \mathbf Y^{k+1}$. 
The episode terminates when all the clients have been visited or no resource remains. 
For an DSCOM, considering the dynamic entering of clients and resource recycles, we set a time interval to maintain the Markov property. 
Within each time interval, the clients enter and wait. 
At each time interval, the DRL agent makes sequential actions as in SCOM and determines whether to recycle the resource based on the time the clients occupy at the last time interval. 
The state transmits to $\rho$, $\mathbf X = \mathbf 0$ and $\mathbf Y= \mathbf 0$ at the next time interval. 
The episode terminates after specific time intervals.

\textbf{Reward.} Our objective is to maximize revenue. 
The reward is designed as the payment $\tau_{ij}$ at each round (i.e., Eqn.~(\ref{tau})).

\subsection{Training Algorithm}

\begin{figure}[!t]
\begin{center}
\setlength{\abovecaptionskip}{0.1cm}
\includegraphics[width=0.8\linewidth]{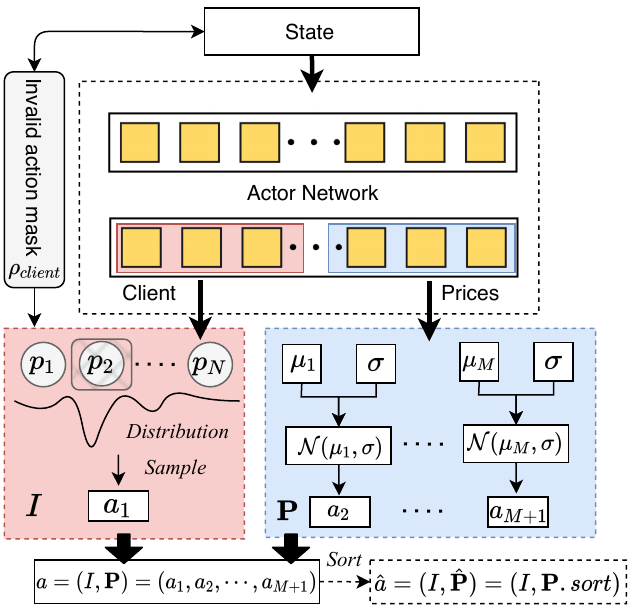}
\caption{Two-part output of the composite action of Egret.}
\label{fig:actions}
\end{center}
\end{figure}

Motivated by its good performance and broad applicability for all \textit{action space} types, we leverage the proximal policy optimization (PPO)\cite{schulman2017proximal} that includes Generalized Advantage Estimation (GAE) \cite{schulman2015high}, clipped surrogate objective, and importance sampling techniques as the training algorithm. 
The actor-critic network is constructed on a multilayer perceptron (MLP) of the same architecture. 
The policy takes the current state as inputs and outputs $N+M$ dimensions continuous values. 
As shown in Fig \ref{fig:actions}, in the last layer of the neural network, the first $N$ elements constitute the partial output of the client index, and the last $M$ elements constitute the prices. 
Specifically, the first $N$ values $p_{1},p_{2},\cdots,p_{N}$, as the probability of each client, will create a categorical distribution to sample the next visiting client $i$ after normalization. 
The last $M$ values, serving as the mean, form $M$ normal distributions to sample the prices $\mathbf P$ the client faced. 
Two parts of the subactions concatenate as the agent's final decision $a=(I, \mathbf P)$.

The training goal is to learn an optimal policy so that the accumulative reward is maximal, which is expressed as
\begin{equation}
\max_{\theta}L(\theta)=\hat{\mathbb{E}}_t\left[\log \pi_\theta\left(a_t \mid s_t\right) \hat{A}_t\right],\label{eq:plain}
\end{equation}
where $\hat{\mathbb{E}}_{t}$ symbolizes a Monte Carlo estimation of the policy gradient. And $\hat{A}_{t}$ is an estimator of the advantage function at timestep $t$. 

Considering the reusability of the sampled trajectories from different distributions, PPO utilizes the importance-sampling technique transferring equation (\ref{eq:plain}) to
\begin{equation}
L^{C}(\theta)=\hat{\mathbb{E}}_t\left[\frac{\pi_\theta\left(a_t \mid s_t\right)}{\pi_{\theta_{\text {old }}}\left(a_t \mid s_t\right)} \hat{A}_t\right]=\hat{\mathbb{E}}_t\left[r_t(\theta) \hat{A}_t\right].\label{eq:is}
\end{equation}
The $r_{t}(\theta)$ is the policy probability ratio of the same trajectory on the new and old policy. 
In order to avoid drastic policy updates, PPO limits the policy gradient by a clipped function so that the old and new policies do not change much and achieve stable training, as
\begin{equation}
L^{C}(\theta)=\hat{\mathbb{E}}_t\left[\min \left(r_t(\theta) \hat{A}_t, \operatorname{clip}\left(r_t(\theta), 1-\epsilon, 1+\epsilon\right) \hat{A}_t\right)\right],
\end{equation}
where $\epsilon$ is a clip coefficient, and it cuts off the part of $r_{t}(\theta)$ that are out of $(1-\epsilon,1+\epsilon)$. 
In a nutshell, the minimum of clipped and unclipped objectives is selected as the final surrogate objective. 
In addition, by adding the value function loss for the critic network and the entropy bonus, the final target of PPO can be expressed as
\begin{equation}
L^{PPO}(\theta)=\hat{\mathbb{E}}_t\left[L_t^{C}(\theta)-c_1 L_t^{V F}(\theta)+c_2 S\left[\pi_\theta\right]\left(s_t\right)\right],\label{eq:ppo}
\end{equation}
where $c_{1}$ and $c_{2}$ are hyper-parameters, $S$ denotes an entropy bonus, and $L^{VF}$ is a squared-error loss between predicted state values $v_{\pi}(s)$ and target state values $\hat v_{\pi}(s)$, as
\begin{equation}
L^{\mathrm{VF}}(\theta)=\hat{\mathbb{E}}_{t}\left[\sum_{t=1}^n\left(v_\pi\left(s_t\right)-\hat{v}_\pi\left(s_t\right)\right)^2\right],
\end{equation}
where 
\begin{equation}
\hat{v}_\pi\left(s_t\right)=\sum_{k=0}^{n-t+1} \gamma^k r_{t+k}. \label{eq:vhat}
\end{equation}
 
\subsection{Enhancement of DRL}\label{TrainEhance}
We employ several techniques to improve the training process and enhance the efficiency of DRL training within the SCOM setting.

\textbf{Price Ranking Technique.} In keeping with the fact that the offered prices must be positively correlated with the resource configuration, the prices must be incremental, i.e., $F_j<F_{j+1}$. 
We implement an action post-processing $\hat {\mathbf  P} \leftarrow \mathbf P.sort()$ for the output prices. 
Consequently, the action becomes $\hat {a_{t}}\leftarrow(I, \hat{\mathbf P})$. 
It is important to note that we still calculate the $\pi_{\theta}(a_{t}\mid s_{t})$ using the original action $a$ so that the post-processing will not affect the training. 
We report the ablation method's performance in section \ref{sec_4} to confirm its necessity.

\textbf{Invalid Action Masking Technique.} It may occur that sampling the client that has been visited or does not exist in this time interval inevitably violates the mechanism. 
The invalid action masking technique is widely employed in the case of blocking some infeasible actions from a discrete space~\cite{vinyals2017starcraft}. 
Its feasibility and effectiveness have been proved~\cite{huang2020closer}. 
Therefore, as shown in Fig \ref{fig:actions}, we adopt the $\rho^{k}_{client}$ as the invalid actions mask at round $k$ so that all those clients who have been visited will not be sampled again.

\textbf{State Space Optimization.} 
We observe that in each interaction, only one row of the allocation matrix $\mathbf X$ and the payment matrix $\mathbf Y$ changes. 
We attempt to reduce the state space by incorporating only the rows that have changed in the allocation and payment matrices into the state at each step.
As a result, the state dimension is reduced to $2\times M$ from $2N\times M$, significantly shrinking the state space. 
We report the improvement of the state optimization on training efficiency and performance in Sec.~\ref{sec:deepdive}. 

\textbf{Generalized Advantage Estimation.}
We use GAE as the advantage function form, which is defined as
\begin{equation}
\hat{A}_{t}=\delta_{t}+(\gamma \lambda) \delta_{t+1}+\cdots+\cdots+(\gamma \lambda)^{T-t+1} \delta_{T-1}\label{eq:gae},
\end{equation}
where $\delta$ denotes the TD-error at time step $t$: $\delta_{t}=r_{t}+\gamma v(s_{t+1})-v(s_{t})$ and $t\in[0,T]$. 
The $\hat A_{t}$ is used to evaluate the quality of the state-action pair, 
and $\hat A_{t} > 0$ means the action $a_{t}$ is relatively good; otherwise relatively poor.
GAE introduces the $\lambda$ to balance the variance and bias of the advantage function.
We discuss the impact of $\lambda$ on training efficiency in section \ref{sec:deepdive}.

\begin{algorithm}[!t]  %
	\caption{Training algorithm of Egret}%
	\LinesNumbered %
	\KwIn{Initialize two neural networks with parameter $\pi_{\theta}(a|s,\theta),\pi_{\theta_{old}}(a|s,\theta)$, experience buffer $\mathbf B$; max step in a trajectory $\mathbf T$; and learning rate $\alpha$.} %
	\KwOut{the Egret's policy $\pi_{\theta}$}%
	$\theta_{old}\leftarrow\theta$\;
	\For {$episode=1,2,\ldots, $}{
	Initialize the state $s_{0}$\;
	\While {$t<\mathbf T$}{\#\textit{Generate a trajectory and auto reset when an episode finishes}\;
	$a_{t}=(I,\mathbf P)\leftarrow$ \textit{actor}$(s_{t},\rho^{t}_{client})$ \#\textit{action mask}\;
	$\hat a_{t}\leftarrow(I,\mathbf P$\textit{.sort()}) \#\textit{pirce ranking technique}\;
	$s_{t+1},r_{t}\leftarrow$\textit{step.}$(\hat a_{t})$
	}
	\For {\textit{each step} $t\in J$}{
	Compute advantage estimates $\hat A_{t}$ according to Eqn.(\ref{eq:gae})\;
	Compute the target state values $\hat{v}_{\pi}(s_{t})$ according to Eqn.(\ref{eq:vhat})\; 
	Store $s_{t}$, $a_{t}$, $\pi_{\theta}(a_{t}\mid s_{t})$, $r_{t}$, $\hat A_{t}$, $v$ and $\hat{v}$ in $\mathbf B$\;
	}
	\For {$epoch=1,2,\ldots$}{
		Randomly sample a minibatch in $\mathbf B$ and compute the PPO target according to Eqn.(\ref{eq:ppo})\;
		Update $\pi_{\theta}$ by SGA with learning rate $\alpha$\;
		$\theta \leftarrow \theta+\alpha \nabla L^{PPO}(\theta)$\;
	}
		Copy new parameters to old: $\theta_{old}\leftarrow\theta$\;
		Clear the experience buffer $\mathbf B$;
	}
\end{algorithm}

Overall, the pseudo-codes of Egret's training are shown in Algorithm 2. 
First, we initialize the updating and target actor-critic networks with the same random parameters (Line 1). 
For each episode, we collect a trajectory of fixed length. 
During the interaction, the agent takes the state $s_{t}$ and action mask $\rho^{t}_{client}$ as inputs and outputs the vanilla action $a_{t}$ at time step $t$. 
After price ranking, agent takes action $\hat a_{t}$ and then obtains the next state $s_{t+1}$ and reward $r_{t}$ (Lines 4-8). 
Next, the advantage estimates $\hat{A_{t}}$ and the estimated state values $\hat{v}_{\pi}$ are calculated and stored in experience buffer $\mathbf B$ and store necessary information, e.g. $s_{t}$, $a_{t}$, and $r_{t}$, etc. (Lines 10-13). 
Eventually, we update the $\pi_{\theta}$ for several epochs. 
For each epoch, the agent optimizes its networks by mini-batch stochastic gradient ascent (SGA) from the buffer (Lines 15-19). 
In addition, the experience buffer should be cleared, and synchronize the old and new parameters (Lines 20-21).

\section{Performance Evaluation}\label{sec_4}
We implement a prototype of Egret with Python. 
We conduct extensive experiments to reveal the performance of Egret for both SCOM and DSCOM settings.

\subsection{Experimental Settings}
The parameters of the client follow the rule that each one has a task of data $d_{i}\sim \mathcal N(3,0.1)Mb$ with local computing capacity drawn from $(1.00,2.00)Mb/s$, and network bandwidth uniformly ranges from $(0.30,0.50)Mb/s$. 
We set the client's unit transmission cost $v = 1e^{-3}$ and energy dissipation coefficient $\mu = 1e^{-2}$~\cite{tran2019federated,dinh2020federated,zhan2020deep}. 
The server holds ten different types of resources with computing capacity $\{5,10,\cdots,50\}Mb/s$. 
Weights are set as $\alpha=0.1$, $\beta=1$, and $\gamma = 1$. 
We consider a set of 5 types of clients for SCOM and 20 for DSCOM.

Ideally, our objective is to train a general agent which can accurately make sequential decisions across any client combination while achieving the maximum revenue in the DSCOM setting. 
To this end, comprehensive arriving traces are generated as training datasets to cover all possible scenarios.
Specifically, we set the length of an episode $l=20$ to mitigate the drastic fluctuation of reward caused by the randomness of short episodes. 
At each discrete step (every 3 seconds), we draw a random number $(\leq 20)$ of clients as the awaiting clients.
We set $1\times10^6$ and $2\times10^7$ total interaction steps for Egret under SCOM and DSCOM, respectively.

The hyperparameters of PPO are tuned by Wandb's Sweep~\cite{wandb}. 
We initiate a pre-training phase on a concise episode to identify the optimal architecture for the actor and critic networks (refer to Sec.~\ref{sec:deepdive}).
Subsequently, the network demonstrating the highest revenue during the pre-training phase is subjected to further training across an extended series of episodes. 
In particular, both the actor and critic networks have three fully connected layers with 64, 64, and 64 neurons, respectively. 
The $\tanh$ activation function is employed after each layer. 
We also apply batch normalization for reward and state for better performance~\cite{pmlr-v48-duan16}. 
We consider the following five baseline methods:

$\bullet$ Oracle (Sec.~\ref{opt}): In scenarios where clients' information is fully observable, the optimal visiting order and prices are unique and deterministic. 
For DSCOM, the ECSP possesses knowledge only of the client in the queue, not the whole trajectory.

$\bullet$ EgretN: It removes the price ranking operation compared to Egret.

$\bullet$ Random Order Policy (ROP): The DRL agent only outputs the prices, while the visiting order is randomly initialized at the beginning.

$\bullet$ Random Price Policy (RPP): The DRL agent only outputs the visiting client, while the prices are randomly generated at each round.

$\bullet$ Deep-SPM~\cite{brero2021reinforcement}: An DRL-based approach trained in the static environment.

\subsection{Experimental Results}
\textbf{Part \uppercase\expandafter{\romannumeral1}: Sequential Computation Offloading Mechanism}

 \begin{figure*}[htbp]
 \begin{center}
 \setlength{\abovecaptionskip}{0.cm}
 \includegraphics[width=1.0\linewidth]{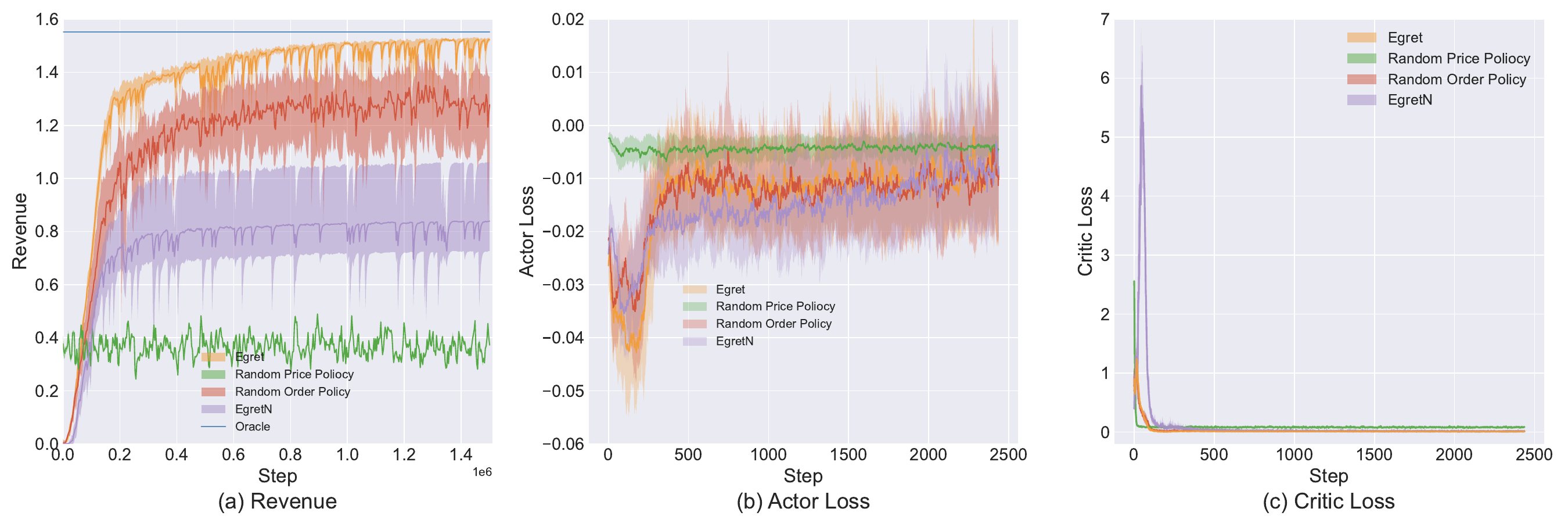}
 \caption{(a) Training episodic return of Egret in SCOM setting. (b) Actor loss and (c) critic loss in the training process. The color strips represent the performance groped of three different seeds.}%
 \label{fg:3}
 \end{center}
 \end{figure*}

We first analyze the revenue performance of the proposed approaches. 
The episodic return and loss during training are depicted in Fig.~\ref{fg:3}. 
The blue line marks the theoretical optimal revenue (i.e., $R = 1.551$). 
We notice that the average reward of Egret increases sharply at the beginning and steadily grows after $2\times10^{5}$ steps, then converges to a steady state in approximately 8000 episodes. 
Our actor-critic loss shows that the training converges after around 500 updates, and the incipient loss peaks are caused by exploration with a high learning rate. 
The highest revenue of Egret reaches 1.536, which is $1.83\times$, $1.09\times$, and $2.79\times$ compared with EgretN, ROP, and RPP, respectively. 
Apparently, the ROP trend is similar to Egret but fails to reach its peak because the visiting order is a necessary component that must be compulsorily controllable to the ECSP. 
EgretN fails due to the misalignment between price and resource, which emphasizes the criticality of the price ranking technique. 
As for RPP, arbitrary price offers always keep it in low revenue. 
Consequently, we deduce that the Egret is not only learnable but also efficacious within the SCOM framework.

\begin{figure}[!t]
\begin{center}
\setlength{\abovecaptionskip}{0.cm}
\includegraphics[width=0.9\linewidth]{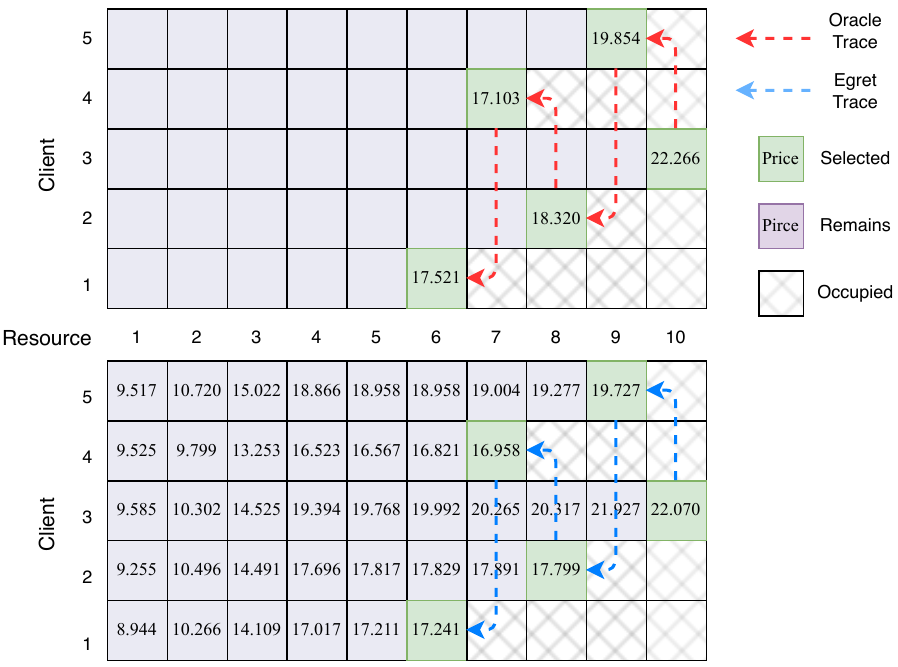}
\caption{The price matrix and the trace comparison between Oracle and Egret with final revenue 1.551 and 1.533, respectively. The arrows denote the visiting order of the edge server. 
The numbers on the box are the prices offered of each round for each resource.}

\label{fig:4}
\end{center}
\end{figure}

\begin{figure*}[htbp]
\begin{center}
\setlength{\abovecaptionskip}{0.cm}
\includegraphics[width=1\linewidth]{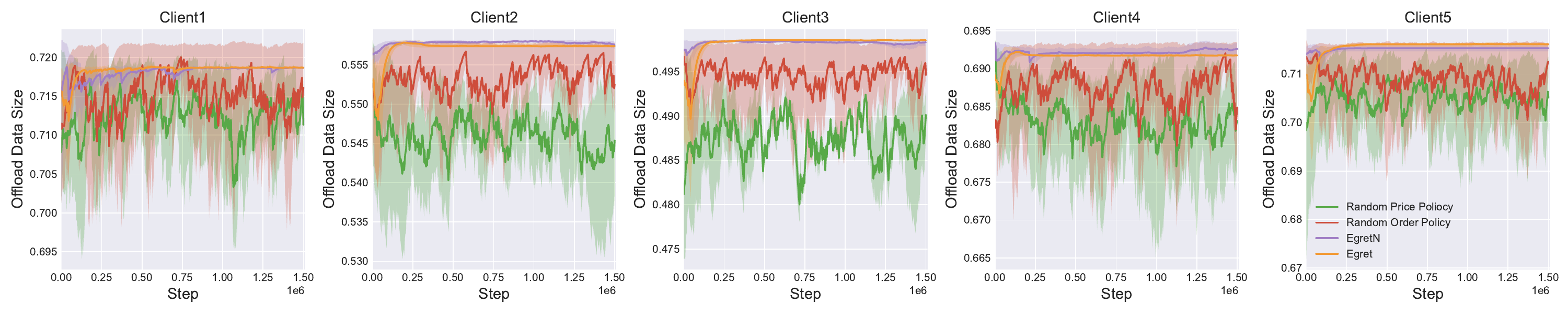}
\caption{The episodic offloading data size of each client.}
\label{datasize}
\end{center}
\vspace{-3mm}
\end{figure*}
\begin{figure*}[htbp]
\begin{center}
\setlength{\abovecaptionskip}{0.cm}
\includegraphics[width=1\linewidth]{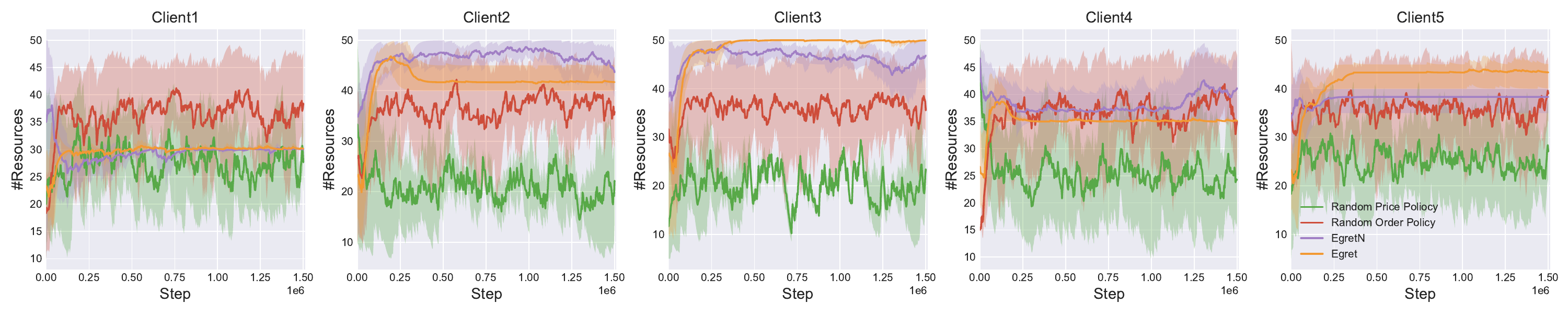}
\caption{The episodic resource that the client chooses to rent.}
\label{res}
\end{center}
\vspace{-3mm}
\end{figure*}
\begin{figure*}[htbp]
\begin{center}
\setlength{\abovecaptionskip}{0.cm}
\includegraphics[width=1\linewidth]{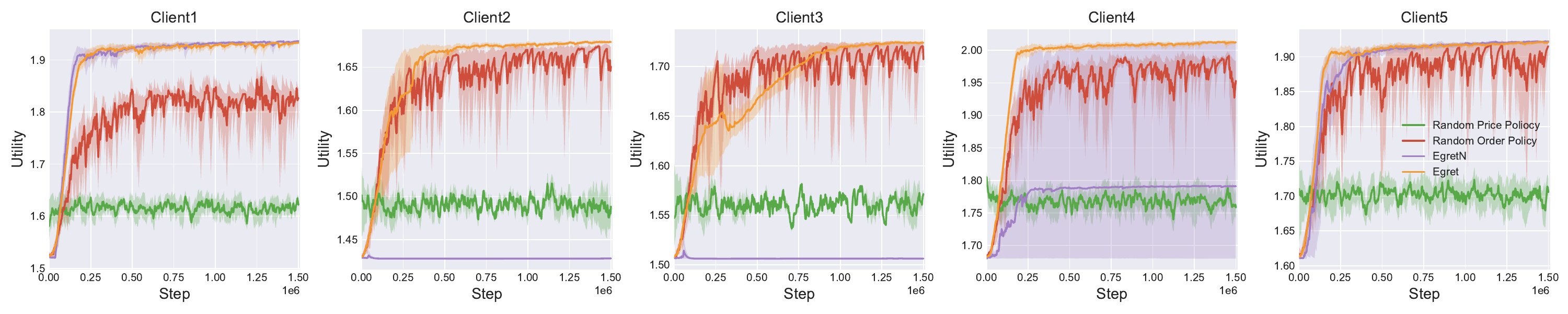}
\caption{The utility of each client.} %
\label{utility}
\end{center}
\vspace{-3mm}
\end{figure*}

Fig.~\ref{datasize},~\ref{res}, and \ref{utility} display the clients' offloading data size, the type of rented resources, and their utility throughout the training phase, respectively.
Our analysis indicates that specific methods, such as ROP and EgretN, can sporadically attain substantial profits. 
However, their post prices or visiting orders tend to fluctuate, leading to notable variances in client decisions. 
In contrast, Egret efficiently explores optimal visiting orders early in the training phases and gradually exploits them by refining resource pricing, resulting in stable individual client utility after $5\times 10^5$ of steps. 
For example, in Fig.~\ref{utility}, the final utilities of the five clients converge to 1.933, 1.679, 1.723, 2.013, and 1.922, respectively.

Fig.~\ref{fig:4} is a concrete example showing the consistency between the Egret and the Oracle. 
The price matrix is the $5\times10$ dimensional matrix, where each row records the posted price for every available resource. 
The arrow signifies the ECSP's visiting order. 
The green box highlights the resource chosen in each round, while the purple boxes represent unselected resources.  
For example, the Oracle trace is $<$3$\rightarrow$5$\rightarrow$2$\rightarrow$4$\rightarrow$1$>$ with price $[22.266, 19.854, 18.320, 17.103, 17.521]$. 
Both approaches prefer to sell the resource with powerful computing capacity first. 
The meticulous decision made by Egret exactly leads to the same visiting order as Oracle's.
Furthermore, Egret's price ranking technique ensures that the prices it posts in each round closely match those set by Oracle.
In summary, Egret can quickly align with the optimal scenario, achieving maximum revenue in the SCOM context.

\textbf{Part \uppercase\expandafter{\romannumeral2}: Dynamic Sequential Computation Offloading Mechanism}

\begin{figure}[!t]
\begin{center}
\setlength{\abovecaptionskip}{0.cm}
\includegraphics[width=0.7\linewidth]{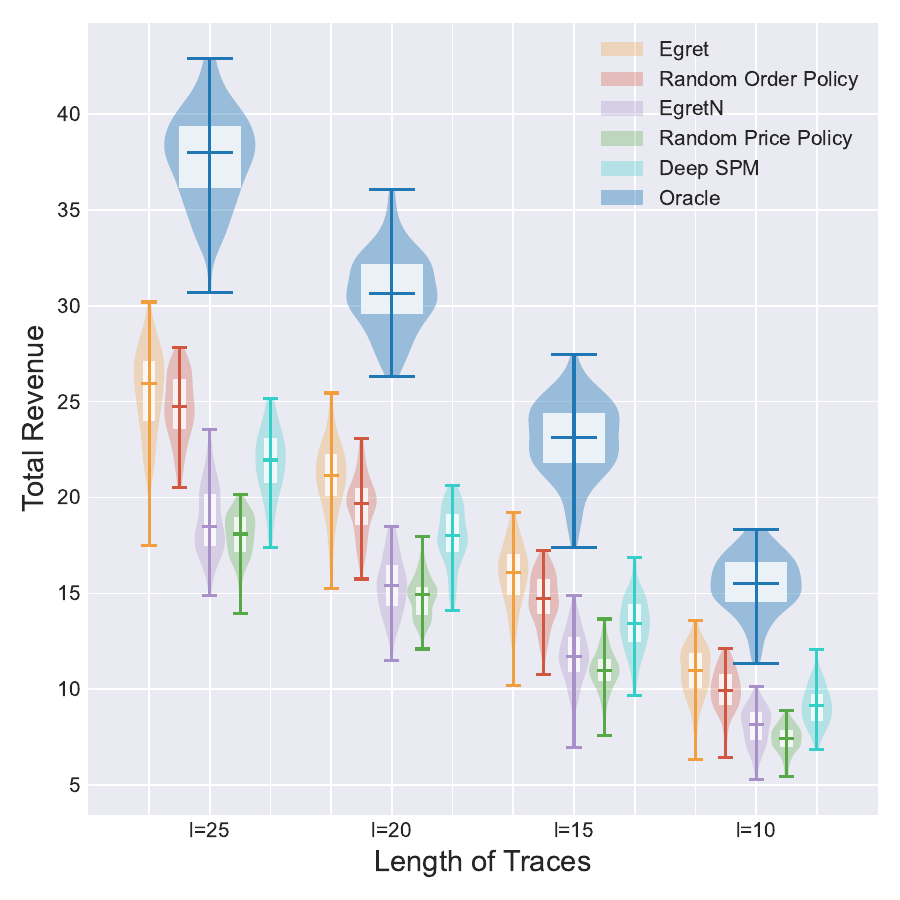}
\caption{Total revenue in different lengths ($l=25,20,15,10$) trace. Shaded areas indicate data distribution, and the white box covers the $25\%-75\%$ percentile of revenue. }
\label{fig:dspm1}
\end{center}
\end{figure}

To further demonstrate the adaptability of Egret within a continuous dynamic environment, we generated a total of 400 traces.
These traces were evenly divided among four different lengths, with 100 traces each for lengths of 10, 15, 20, and 25, respectively.
An arbitrary number of clients enter at every time interval. 
The agent determines the advantageous visiting order and prices for each time interval only based on current information. 
The performance of Egret in the DSCOM setting is shown in Fig.~\ref{fig:dspm1}. 
The violin plot shows the revenue distribution, including the upper and lower bound, the median number, and the white box covers the $25\%-75\%$ percentile of revenue. 
Egret outperforms other baselines and exhibits minimal error from the Oracle solution.
In contrast, EgretN cannot send the posted prices to the correct client, thereby impairing the client's choice. 
RPP, on the other hand, even receives little revenue due to the excessive uncertainty in its pricing.
Deep-SPM was trained solely on static data, simulating a completely inexperienced agent. 
The findings reveal its inability to provide suitable prices or determine visiting orders that adapt to fluctuating client preferences.
However, as the length of an episode extends, the revenue margin between these methods and the Oracle solution progressively expands, a result of the error that inevitably accumulates with each step.

Table~\ref{tab2} details the average revenue margin and margin per step across various approaches compared to the Oracle solution. 
We highlight the average margin error for a clearer comparison.
Egret consistently maintains a low error level of approximately $0.48$ across all lengths, whereas the errors for EgretN, ROP, RPP, and Deep-SPM are about $1.61\times, 1.20\times$, $1.76\times$, and $1.39\times$ higher than Egret's, respectively.
Overall, Egret's average error diverges from the Oracle by merely 31.22\%.
These experimental outcomes validate that our DRL approach has successfully learned a specific strategy for handling a group of clients.

\begin{table}[!t]
\setlength{\abovecaptionskip}{0cm}  %
\setlength{\belowcaptionskip}{0cm} %
\renewcommand{\arraystretch}{1.3}
\centering
\caption{Average revenue margin and margin per step in different trace lengths.}
\label{tab2}
\resizebox{\columnwidth}{!}{%
\begin{tabular}{@{}c|c|c|c|c|c@{}}
\toprule
\textbf{} & \textbf{10} & \textbf{15} & \textbf{20} & \textbf{25} & \textbf{Avg. Margin} \\ \midrule
\textbf{Egret}                 & 4.63/\textbf{0.46}  & 7.12/\textbf{0.47}  & 9.72/\textbf{0.49}  &  12.22/\textbf{0.49}   & \textbf{31.22\%}        \\
\textbf{EgretN}                & 7.44/\textbf{0.74}  & 11.28/\textbf{0.75}  & 15.31/\textbf{0.77}  &  18.91/\textbf{0.76}   & \textbf{49.26\%}        \\
\textbf{Random Order}          & 5.55/\textbf{0.55}  & 8.31/\textbf{0.55}  & 11.10/\textbf{0.55}  &  13.00/\textbf{0.52}   & \textbf{35.63\%}        \\
\textbf{Random Price}          & 8.11/\textbf{0.81}  & 12.03/\textbf{0.80}  & 15.91/\textbf{0.80} &  19.75/\textbf{0.79}   & \textbf{52.20\%}         \\ 
\textbf{Deep-SPM}          & 6.36/\textbf{0.64}  & 9.65/\textbf{0.64}  & 12.77/\textbf{0.64} &  15.90/\textbf{0.64}   & \textbf{41.68\%}         \\ 
\bottomrule
\end{tabular}%
}
\end{table}

\begin{figure*}[htbp]
\begin{center}
\setlength{\abovecaptionskip}{0.cm}
\includegraphics[width=1\linewidth]{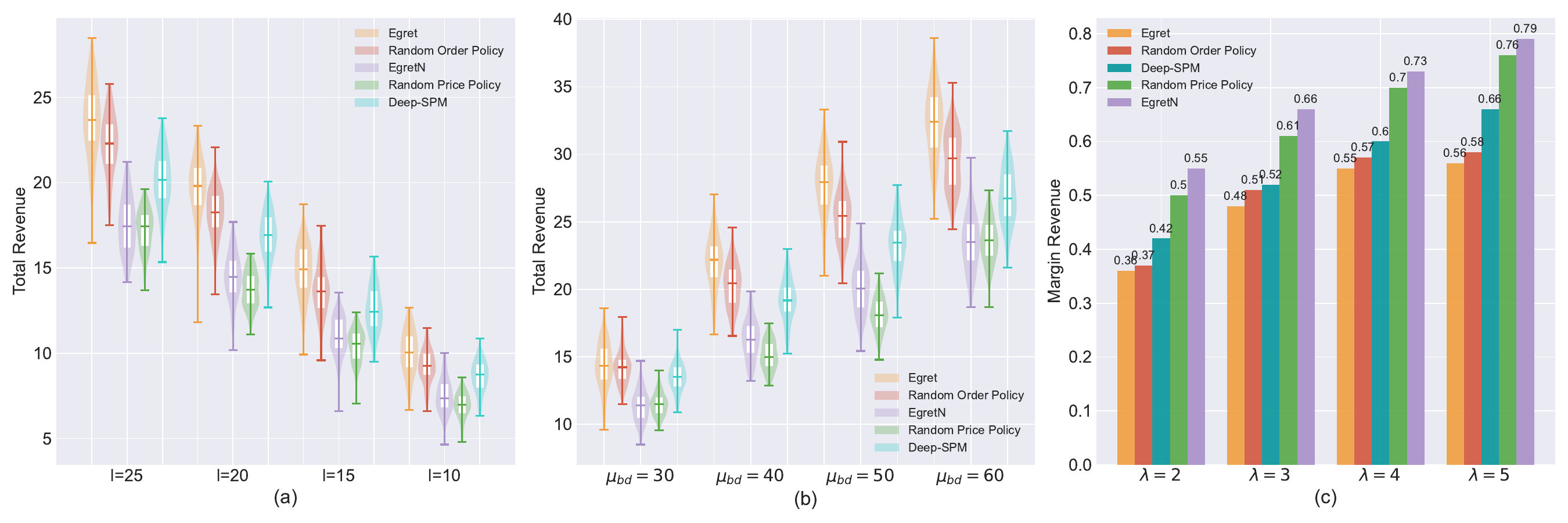}
\caption{(a) Total revenue when the clients with fluctuating input data. (b) Influence of varying bandwidth on revenue. (c) Average margin per-step revenue of different clients amount. The number of clients that enter the environment at each time interval satisfies the Poisson distribution $\lambda =2,3,4$, and $5$.}
\label{fig:7}
\end{center}
\end{figure*}

Moreover, we evaluate the performance of various methods amidst variations in client computing tasks.
We adjust the input data to satisfy $\mathcal N(\text{unif}(2,5),0.5)Mb$ whenever a new client enters. 
Fig.~\ref{fig:7}(a) demonstrates the resilience of Egret against fluctuating input data across various trace lengths.
After training on a range of trace data, Egret and ROP managed to devise a more effective strategy, although slightly inferior to the previous static input data. 
Conversely, Deep-SPM, reliant on static trace data, struggles to adapt to changing conditions and underperforms compared to ROP.
In summary, Egret significantly exceeds others by having the highest average revenue. 
Specifically, the average revenue of Egret with 25 length is 27.63, which is about $1.04\times$, $1.44\times$, $1.34\times$, and $1.21\times$ more than ROP, RPP, EgretN, and Deep-SPM respectively.

Subsequently, we set the trace length to $20$ and adjust the bandwidth for each client by sampling from $\frac{\mathcal N(\mu_{bd},5)}{100} Mb/s$.
Fig.~\ref{fig:7}(b) shows the average revenue when $\mu_{bd}=30$, $40$, $50$ and $60 Mb/s$. 
As bandwidth gradually increases, the transmission time for clients decreases, eliminating the limitation on selecting data offloading volume. 
According to the analysis in Sec.~\ref{opt}, the server can suitably elevate the pricing to ensure the client's utility reaches the initial baseline, thus attaining a higher revenue. 
Overall, Egret exhibits superior performance by achieving revenue up to $1.38\times$, $1.09\times$, $1.52\times$, and $1.20\times$ compared with EgretN, ROP, RPP, and Deep-SPM, respectively.

Last, we evaluate the performance under harsh conditions. 
We assume the number of clients entering the environment at each time interval follows a Poisson distribution, and we adjust the parameter $\lambda$ from 2 to 5 while maintaining a constant total of 20 steps to simulate increased pressure.
\begin{equation}
P(X=k)=\frac{\lambda^k}{k !} e^{-\lambda}, k=0,1, \cdots
\end{equation}
Fig.~\ref{fig:7}(c) depicts the margin per step of each approach. 
The margin gets larger when the number of users entering each step becomes larger, but Egret keeps a slight increase compared with the baselines. 
Overall, Egret shows high adaptability by shrinking the margin up to $1.53\times$, $1.06\times$, $1.39\times$, and $1.18\times$ compared with EgretN, ROP, RPP, and Deep-SPM, implicating that Egret has high robustness when facing pressure, and learns a similar strategy as the Oracle.

\subsection{Deep Dive into the Egret}\label{sec:deepdive}

We finally evaluate our detailed design of Egret.

\begin{figure}[!t]
\begin{center}
\setlength{\abovecaptionskip}{0.cm}
\includegraphics[width=1\linewidth]{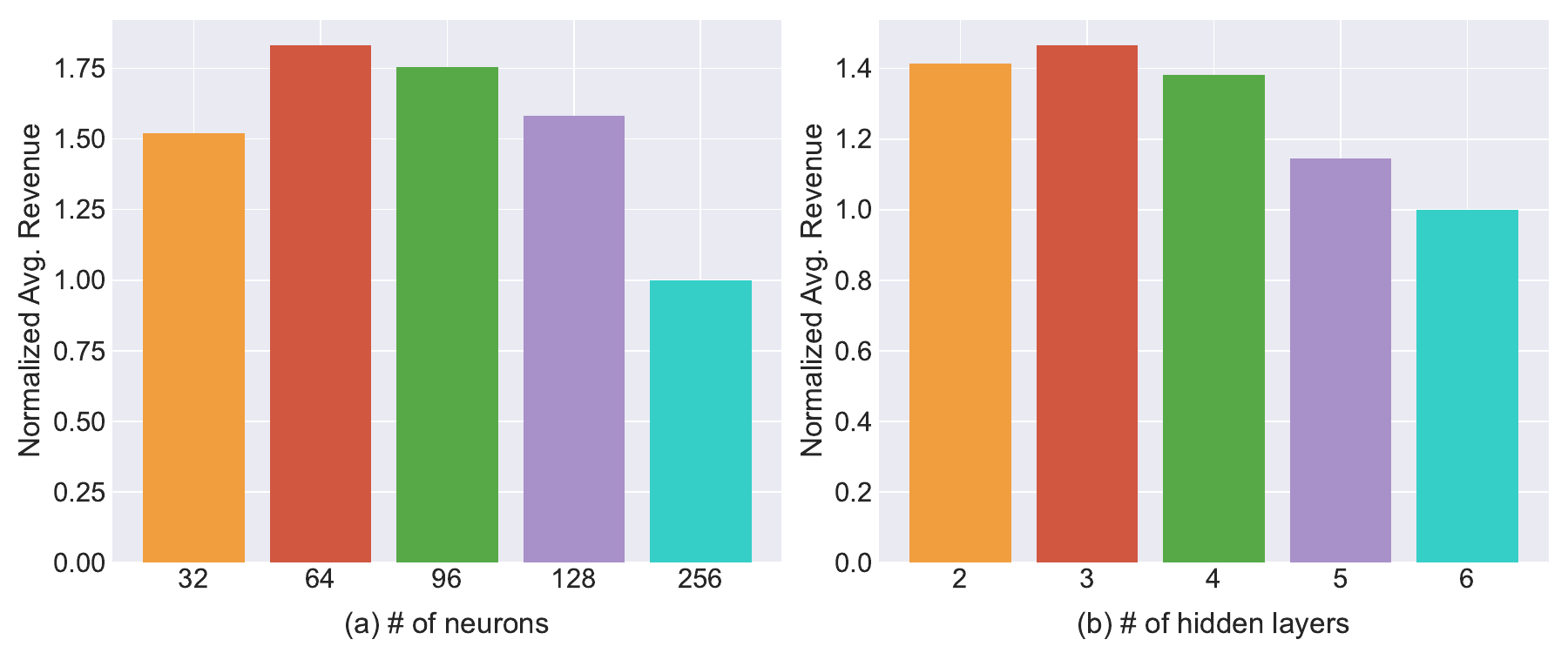}
\caption{Normalized average revenue with varying structures}
\label{fig:deep}
\end{center}
\end{figure}

\textbf{Number of Neurons.}
For both actor and critic networks, we maintain a constant three hidden layers while altering the neuron count in each hidden layer, ranging from 32 to 256.
All neural networks undergo training on the identical dataset encompassing $2\times10^{7}$ trajectory steps and are evaluated on the same episodes.
Fig.~\ref{fig:deep}(a) shows the normalized performance on different networks, with optimal results observed at 64 neurons.

\textbf{Number of Hidden Layers.}
We set the number of neurons to 64 while altering the depth of hidden layers from 2 to 6. 
As depicted in Fig.~\ref{fig:deep}(b), there is no obvious linear correlation between the number of hidden layers and the revenue. 
Networks consisting of three layers marginally outperform their counterparts.
The performance of the first three neural networks is nearly the same, but with more neural network layers, it generally deteriorates due to the gradient vanishing problem.

\textbf{Lambda in Generalized Advantage Estimation.}
Table~\ref{tab3} presents the ratio of the reward to the Oracle (i.e., reward/optimal value) for Egret and its variants within the SCOM setting over a training duration of $10^6$ steps.
Each reported value represents the mean derived from trials using three random seeds.
The experiments explored adjustments to the GAR $\lambda$ parameter, including setting it to 0 (equivalent to TD(0) advantage estimation), setting it to 1 (equivalent to Monte Carlo advantage estimation), and excluding GAE in favor of employing discounted reward-to-go. 
The experimental results show that Egret~($\lambda = 0.95$) can obtain the maximum learning potential and has a fast convergence speed.

\begin{table}[!t]
\setlength{\abovecaptionskip}{0cm}  %
\setlength{\belowcaptionskip}{0cm} %
\centering
\caption{The Ratio of Reward to Oracle}
\scalebox{1}{
\label{tab3}
\begin{tabular}{l|ccccc}
\hline
\multicolumn{1}{c|}{}                          & \multicolumn{5}{c}{\textbf{Percentage of the training process. (\%)}} \\ \hline
\multicolumn{1}{c|}{\textbf{Methods}} & \textbf{5}  & \textbf{10} & \textbf{15} & \textbf{20} & \textbf{25}  \\ \hline
\multicolumn{1}{l|}{Egret ($\lambda=0.95$)}    & 18.56    & 48.02    & \underline{\textbf{79.00}}    & \underline{\textbf{87.09}}    & \underline{\textbf{89.46}}   \\
\multicolumn{1}{l|}{B-Egret ($\lambda=0.95$)}  & 19.33    & 36.20     & 49.70    & 50.78    & 53.30   \\
\multicolumn{1}{l|}{Egret-MC ($\lambda=1$)}    & 16.07    & 39.58     & 67.48   & 77.99    &         79.82 \\
\multicolumn{1}{l|}{Egret-TD(0) ($\lambda=0$)} & \underline{\textbf{20.73}}    & \underline{\textbf{55.98}}    & 77.39    & 84.35    & 86.19   \\
\multicolumn{1}{l|}{Egret-w/oGAE}             & 16.07    & 36.64    & 68.82    & 78.53    & 81.60   \\ \hline
\multicolumn{1}{c|}{\textbf{Methods}}          & \textbf{30} & \textbf{40} & \textbf{60} & \textbf{80} & \textbf{100} \\ \hline
\multicolumn{1}{l|}{Egret ($\lambda=0.95$)}    & \underline{\textbf{88.89}}    & \underline{\textbf{92.41}}    & \underline{\textbf{95.11}}    & \underline{\textbf{96.46}}    & \underline{\textbf{98.17}}   \\
\multicolumn{1}{l|}{B-Egret ($\lambda=0.95$)}  & 54.58    & 54.60    & 55.44    & 55.51    & 55.40   \\
\multicolumn{1}{l|}{Egret-MC ($\lambda=1$)}    & 82.75    & 84.60    & 86.27    & 87.86         & 88.77         \\
\multicolumn{1}{l|}{Egret-TD(0) ($\lambda=0$)} & 87.00    & 89.35    & 91.92    & 92.75    & 93.49   \\
\multicolumn{1}{l|}{Egret-w/oGAE}             & 85.09    & 84.88    & 87.28    & 87.75    & 88.22   \\ \hline
\end{tabular}
}
\end{table}

\textbf{Performance of State Space Optimization}
B-Egret in Table~\ref{tab3} denotes feeding the entire allocation matrix and payment matrix as the state input to the network. 
Experiments indicate that such a complex and redundant state does not enhance performance but rather constrains the capability of policy learning.

\section{Related Works}\label{sec_5}
\textbf{Sequential Price Mechanism.} A series of learning-based algorithms~\cite{huo2022learning,brero2021reinforcement,brero2021learning} were proposed to solve the sequential price mechanism. 
Brero et al. first introduce the RL method to solve the SPM in economic scenarios where there are agents and indivisible items~\cite{brero2021reinforcement}. 
Specifically, Each agent has a valuation function for each item, and an economic mechanism interacts with agents and offers prices for each item. 
The design goal of the mechanism is to maximize social welfare or revenue. 
The main principle is that the RL mechanism can access agents' distribution through interaction and achieve the objective despite incomplete information about agents. 
Furthermore, Brero et al. establish a Stackelberg game model in SPM with small modifications~\cite{brero2021learning}, where the mechanism as the leader takes action first based on its strategy of maximizing social welfare or revenue. 
At the same time, the agents, as the followers, respond to maximize their utilities after that. 
They transfer the multi-agent mechanisms to a single-agent problem through the Stackelberg MDP, then achieve the Stackelberg Equilibrium by RL.

\textbf{Task Offloading in Mobile Edge Computing Systems.}
Previous surveys~\cite{zamzam2020game,feng2022computation} have interpreted the achievements in computation offloading in detail from different approaches and perspectives, including the game-theoretic approach. 
To maximize the revenue, \cite{kim2018optimal,liu2017price} model the interaction between the edge cloud as a Stackelberg game. 
In \cite{zhan2020deep}, Zhan et al. formulated the multi-user edge offloading environment without shared information as a partially observable Markov decision process and then used reinforcement learning to achieve the equilibrium of each user's utility. 
A similar multi-user and multi-server multi-access edge computing (MEC) environment with the goal of maximizing the MEC servers’ profit was established in \cite{li2021deep}. 
They used a two-step optimization method to solve the problem without considering the server's prices. 
Tang et al. use DRL to determine offloading decisions to minimize the expected long-term cost~\cite{9253665}. 
However, it does not involve monetary transactions between the server and the client.
In the dynamic mobile cloud computing environment, \cite{7586102} introduced a new stochastic control algorithm that makes online decisions on computing service purchasing and resource allocation to maximize profit.
\cite{8385143} proposed a multi-agent stochastic learning algorithm to reach the Nash Equilibrium. 

Unlike previous studies, our approach solves the Sequential Price Mechanism problem under a dynamic environment without information sharing, where mobile users can dynamically become active or inactive without sharing information, and it can almost attain the optimal solution.

\section{Conclusion}\label{sec_6}
In this paper, we propose a novel sequential computation offloading mechanism in edge computing. 
We give the theoretical optimum for determining the resource prices and the client visiting orders to maximize the revenue of the ECSP in an Oracle scenario. 
We develop an DRL approach named Egret, with a price ranking technique that achieves near-optimal revenue depending on the client's label and transaction record. 
Our numerical experiments show that Egret surpasses other baseline strategies. Egret's optimal result in the SCOM setting is only 1.289\% lower than the oracle and 23.43\% better than the state-of-the-art in a dynamic SCOM setting.

\ifCLASSOPTIONcaptionsoff
  \newpage
\fi

\bibliographystyle{IEEEtran}
\bibliography{raytheon}

\end{document}